\documentclass[]{emulateapj}
\def\pppm{\rm P^3M}
\def\mpc{\,h^{-1}{\rm {Mpc}}}
\def\kpc{\,h^{-1}{\rm {kpc}}}
\def\kms{\,{\rm {km\, s^{-1}}}}
\def\msun{{\,h^{-1}M_\odot}}
\def\etl{et al.}

\begin{document}
\shorttitle{Dependence of Occupation of Galaxies on Halo Formation Time}
\shortauthors{Zhu \etl}
\title {
The Dependence of the Occupation of Galaxies on the Halo Formation Time}

\author{
 Guangtun Zhu\altaffilmark{1,2},
 Zheng Zheng\altaffilmark{3,4},
 W.P. Lin\altaffilmark{1},
 Y.P. Jing\altaffilmark{1}, 
 Xi Kang\altaffilmark{1,5}, and
 Liang Gao\altaffilmark{6}
} 
\altaffiltext{1}{
Shanghai Astronomical Observatory, Joint Institute for Galaxy
  and Cosmology (JOINGC) of SHAO and USTC, Nandan Road 80, 
Shanghai, 200030, China; gtzhu@shao.ac.cn.}
\altaffiltext{2}{
Graduate School of the Chinese Academy of Sciences, 19A, 
Yuquan Road, Beijing, China.}
\altaffiltext{3}{
Institute for Advanced Study, School of Natural Sciences, 
Einstein Drive, Princeton, NJ 08540.}
\altaffiltext{4}{
Hubble Fellow.}
\altaffiltext{5}{
Astrophysics, University of Oxford, Keble Road, Oxford, OX1, 3RH, UK.}
\altaffiltext{6}{
Institute for Computational Cosmology, University of Durham, DHE 3LE, UK.}

\begin{abstract}
We study the dependence of the galaxy contents within halos on the halo
formation time using two galaxy formation models, one being a
semianalytic model utilizing the halo assembly history from a high
resolution {\it N}-body simulation and the other being a smoothed particle
hydrodynamics simulation including radiative cooling, star
formation, and energy feedback from galactic winds. 
We confirm the finding by Gao et al. that at
fixed mass, the clustering of halos depends on the halo formation time,
especially for low-mass halos. This age dependence of halo clustering
makes it desirable to study the correlation between the occupation of
galaxies within halos and the halo age. We find that, in halos of
fixed mass, the number of satellite galaxies has a strong dependence
on halo age, with fewer satellites in older halos. The youngest
one-third of the halos can have an order of magnitude more satellites
than the oldest one-third. For central galaxies, in halos that form
earlier, they tend to have more stars and thus appear to be more
luminous, and the dependence of their luminosity on halo age is not as
strong as that of stellar mass. The results can be understood through
the star formation history in halos and the merging of satellites onto
central galaxies. The age dependence of the galaxy contents within halos
would constitute an important ingredient in a more accurate halo-based
model of galaxy clustering.
\end{abstract}


\section {Introduction} 

In the cold dark matter hierarchical model of structure formation,
the formation and evolution of galaxies are coupled with those of 
dark matter halos. Studying the distribution of galaxies inside halos 
can aid us in the understanding of the galaxy formation process and in the 
interpretation of clustering properties of galaxies. In this Letter, 
we investigate the dependence of the galaxy contents on the halo formation 
time using numerical galaxy formation models. 

Our study is closely related to recently developed models of galaxy
clustering, namely, the framework of the halo occupation distribution
(HOD) and the approach of the conditional luminosity function
(CLF). Both models are based on statistical descriptions of the
relation between galaxies and dark matter halos. The HOD characterizes
the relation in terms of the probability distribution $P(N|M)$
that a halo of mass $M$ hosts $N$ galaxies of a given type, and the
spatial and velocity distributions of galaxies within halos
(\citealt{Jing98a,Ma00,Peacock00,Seljak00,Scoccimarro01,Berlind02,Cooray02}).
Instead of $P(N|M)$ for galaxies of a given type, the CLF method uses
the luminosity function of galaxies within halos as a function of halo
mass to establish the relation between galaxies and dark matter halos
\citep{Yang03}. HOD and CLF models have
been applied to interpret the observed galaxy clustering in a number
of surveys (e.g.,
\citealt{Jing98a,Jing02a,Bullock02,Moustakas02,Yang03,vandenB03,
Magliocchetti03,Yan03,Zheng04,Zehavi04,Zehavi05,Ouchi05,Lee05,Cooray05}),
and with the help of these models our understanding of galaxy
clustering has been greatly improved and refined.

One key assumption in these halo-based models is that the {\it statistical 
distribution} of galaxies inside halos depends only on halo mass. This 
assumption has its theoretical origin in the excursion set formalism 
\citep{Bond91}, which predicts that a halo's assembly history depends 
only on its mass. It is inherent in any semianalytic galaxy formation 
model that uses halo merger trees based on the excursion set formalism. 
Using the high-resolution Millennium $N$-body simulation, 
\citet{Gao05} (also see \citealt{Harker05}) show that the 
clustering of halos at a given 
mass has a dependence on the halo formation time, and the dependence 
becomes much stronger at halo masses $\ll M_{\ast}$, the nonlinear 
characteristic mass. Galaxy
properties are expected to be correlated with the halo formation time,
and to this extent, the environmental dependence of the halo formation 
time would lead to a dependence of the statistical distribution 
of galaxies inside halos on the environment as well as on halo mass. 
Once high accuracy modeling eventually becomes demanded by galaxy 
clustering data and/or by cosmological applications, the current 
HOD and CLF models may have to be modified to take into account the 
environmental dependence. Given the age dependence of the halo clustering,
the study of galaxy contents as a function of halo age at fixed halo 
mass would shed light on the extension of the HOD/CLF model. Moreover,
the dependence of galaxy properties on the halo formation time 
is an interesting problem on its own, since it can aid us in our understanding 
of the galaxy formation process.

In this Letter, we examine the dependence of the halo occupation 
properties of galaxies on the halo formation time, using a semianalytic
galaxy formation model (SAM) and a smoothed particle hydrodynamics model 
(HYD).
In \S~2, we introduce 
the simulations and galaxy formation models. 
We show a confirmation of the  
age dependence of halo clustering in \S~3.  The dependence 
of galaxy properties on the halo formation time is the content of \S~4,
where we present our results in terms of the CLF as a function of 
halo mass and formation time. In \S~5, we give a summary and discussion.

\section{The Simulations and Galaxy Formation Models}

Two main simulations are used in our study. The one that the SAM is based on
is a $\pppm$ $N$-body simulation in \citet{Jing02} with $512^3$ 
particles in a cubic box of 100 $\mpc$.
A spatially-flat $\Lambda$CDM 
model is adopted with density parameters $\Omega_m=0.3$ and 
$\Omega_\Lambda=0.7$. 
We adopt the cold dark matter power spectrum of \citet{Bardeen86} with
a shape parameter $\Gamma=\Omega_mh=0.2$ and the amplitude $\sigma_8=0.9$.
The softening length is 
$10 \kpc$, and the particle mass is 
$M_p \sim 6.2 \times 10^{8}\msun$. We denote this simulation as 
SAM-$\Lambda$CDM. The galaxies of the SAM sample are generated by 
the model of \citet{Kang05} that explicitly follows
the assembly history of each halo in the simulation.
The model successfully explains a wide variety of observational facts, 
including luminosity functions in different bands and the bimodal color 
distribution of galaxies. 

The other simulation, denoted as HYD-Gadget2, is performed as smoothed
particle hydrodynamics with the TREE-PM code Gadget2
\citep{Springel05,Springel01}, following the evolution of $512^3$ dark
matter particles and $512^3$ gas particles in a cubic box of 100
$\mpc$.  It includes the physical processes of radiative cooling, 
star formation, and energy feedback from galactic winds \citep{sh03}.  
The
cosmological and initial density fluctuation parameters of the model
are ($\Omega_m,\Omega_\Lambda,\Omega_b,\sigma_8,n,h$)=($0.268, 0.732,
0.044,0.85,1,0.71$),
and the transfer function is computed from CMBFAST 
\citep{Seljak96}. 
The softening length is also $10 \kpc$,
and the particle mass (dark matter+gas) is $M_p\sim 5.54
\times 10^{8} \msun$.  Galaxies are identified as gravitationally
bound groups of stars and cold gas particles that are associated with
a common local maximum in the baryon density. We note that the effect
of dust extinction, which is included in the SAM model, is not taken into
account in this model.

In addition to the two main simulations, we also use a low-resolution 
simulation of $512^3$ particles in a box of 300$\mpc$ only for the purpose
of probing the age dependence of halo clustering at high halo mass. 
This simulation (denoted as 300-$\Lambda$CDM) has the same cosmological 
parameters as SAM-$\Lambda$CDM, and the particle mass is $M_p \sim 1.67 
\times 10^{10} \msun$.

In all the simulations, halos are identified with the friends-of-friends 
algorithm \citep{Davis85} with a linking length $0.2$ times the 
mean particle separation. 

\section{The Definition of Formation Time and The Age Dependence of Halo Clustering}
 
\begin{figure}
\epsscale{0.9} 
\plotone{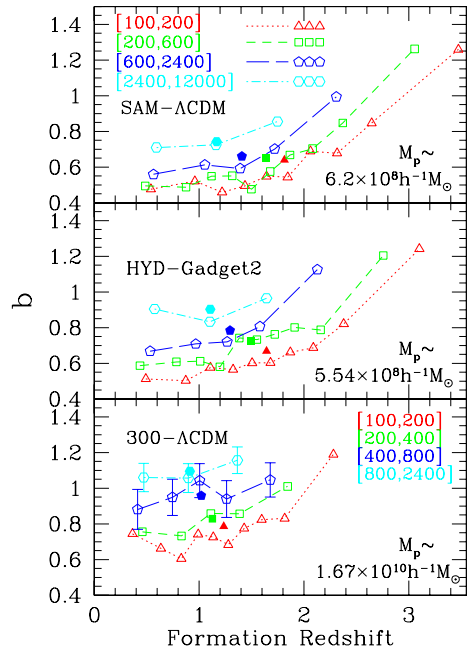}
\caption{
Halo bias factor as a function of formation time.
The two integers in brackets indicate the range of particle numbers of
halos (symbols have the same meaning in the two top panels).
The particle mass is labeled as $M_p$ in each panel.
The open symbols connected by lines are the results in different age bins,
while the filled symbols show the mean formation redshift and bias
factor for all halos in the given mass bin. Poisson error bars are plotted
for the two high-mass bins in the bottom panel.
}
\label{fig1}
\end{figure}

We define the halo formation time as the time when its most massive
progenitor has exactly half of the final mass, interpolated between
two adjacent outputs if necessary. The progenitor of the 
final halo is defined as a halo in an earlier output that has more than
half of its particles found in the final halo. If the most massive
progenitor in an output has more than twice the mass of the one in the
earlier adjacent output, we excluded the 
final halo in our study
since its formation time is difficult to measure because its main
progenitor was likely to be bridge-connected to a massive halo. In the
end, about 10\% of the halos are excluded.

We divide halos in each halo mass bin into several(3, 5, or 10) age bins with an equal
number of halos in each age bin.
The square of the halo bias factor is 
calculated by taking the ratio of the two-point correlation 
functions of halos and matter, averaged over scales $5\mpc < r < 20\mpc$. 

\begin{figure*}
\epsscale{1.1}
\plotone{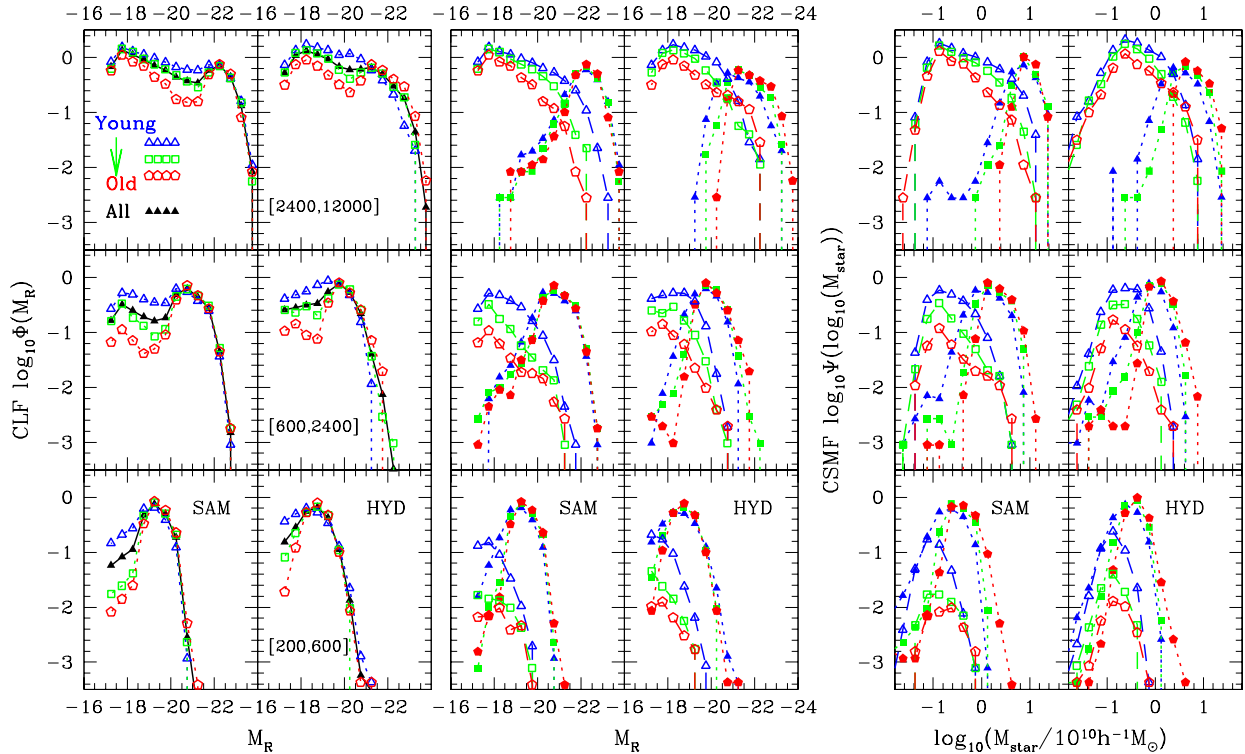}
\caption{
Conditional luminosity function (CLF) and conditional stellar mass
function (CSMF) as a function of halo age. {\it Left two columns}: CLFs
from the SAM ({\it left}) and HYD ({\it right}) galaxy formation models. Results are
shown in three mass bins indicated by the particle numbers [the particle
mass in the SAM (HYD) model is $M_p=$6.2(5.5)$\times 10^8\msun$].
{\it Central two columns}: Same as the left two columns, but the CLF
is separated into contributions from central galaxies ({\it filled symbols}) and
satellites ({\it open symbols}). {\it Right two columns}: Same as the central two
columns, but for the CSMF. The CSMF is not complete at low stellar mass,
which causes the steep drop at the low stellar mass end.
}
\label{fig2}
\end{figure*}

The results of the dependence of halo clustering on the formation time
are shown in Figure~\ref{fig1}. We confirm the finding by
\citet{Gao05} that old halos cluster more strongly than young ones. In
particular, at $\sim 10^{11} \msun$, the large-scale autocorrelation
amplitude for the oldest 10\% of the halos is more than 5 times that for the
youngest 10\% of the halos. For massive halos, the dependence becomes weak.
\citet{Wechsler05} find that the age dependence of the halo clustering
reverses its trend for halos more massive than the nonlinear mass
$M_*$ ($10^{13}$ and $4.6\times 10^{12}\msun$ for the SAM-$\Lambda$CDM
and HYD-Gadget2 simulations in our study), if halo concentration is
used as an approximation of age. We do not detect such a clear trend,
which could be buried in the large (Poisson) uncertainty as indicated
in Figure~\ref{fig1} for the two high-mass bins.

\section{The Age Dependence of The Conditional Luminosity Function}

In this section, we study the dependence of the occupation of galaxies
on the halo formation time from the SAM and HYD galaxy formation models. 
The results are presented in terms of the CLF $\Phi(L|M)$, which is the 
expected number of galaxies per unit luminosity in a halo of mass $M$ 
\citep{Yang03}. Only galaxies brighter than $M_R=-17.0$ are considered 
in our study, where $M_R$ is the $R$-band absolute magnitude. 

The $R$-band CLFs from the SAM and HYD models are shown in the left
two columns of Figure~\ref{fig2}. The drop below $M_R\sim-18$ indicates
that the CLFs are only complete for $M_R<-18$. 
For clarity, halos in each mass bin
are only divided into three age bins with an equal number of halos in
each age bin.  A clear dependence of the CLF on halo age is found in
both models, and the dependence becomes stronger at the faint end of
the CLF.  In general, there are more faint galaxies and fewer bright
ones in younger halos. At the faint end, the youngest one-third of
halos could host 10 times more galaxies than the oldest one-third. At
the bright end, the trend of more bright galaxies in older halos is
clear in the HYD model in the two high-mass bins, and it tends to
disappear at low halo mass ($\sim 10^{11} \msun$). For the SAM model,
the trend at the bright end is not apparent. The subtle difference
between the SAM and HYD models at the bright end could be 
due to the facts that the dust extinction effect is not included in the HYD
model and that the gas cooling is switched off for halos of circular
velocity larger than 350 $\kms$ in the SAM model.

In the central two columns of Figure~\ref{fig2}, the CLFs are separated 
into contributions from central and satellite galaxies, as is done in 
\citet{Zheng05}. At the bright 
end, the CLF is dominated by the bump caused by central galaxies. 
In older halos, central galaxies appear to be brighter, and the central
galaxy bump as a whole shifts to higher luminosity. There are two reasons
for central galaxies being brighter in older halos. First, in older halos, 
star formation is triggered at an earlier time and may last longer. Second, 
the accretion of satellite galaxies into older halos happens earlier, and 
there is enough time for them to merge onto central galaxies. 
Both scenarios increase the amount of stellar components in the central galaxies 
of older halos.
To show that this is indeed the case, we
plot the conditional stellar mass function (CSMF) $\Psi(M_{\rm star}|M)$ in
the right two columns of Figure~\ref{fig2}, separating into contributions
from central and satellite galaxies. 
The CSMF is not complete at the low stellar mass
end ($M_{\rm star}<10^9\msun$), 
which explains the steep drop at this end. It can be clearly seen
that central galaxies in older halos have more stars. Young stellar 
populations are more luminous than old ones, and this reduces, but not 
completely erases, the luminosity difference in central galaxies of young 
and old halos. As a result, the difference in the central galaxy CLFs
in halos of different ages is not as large as that in the central galaxy CSMFs.
In the lowest mass bin, the HYD model may have more stars formed 
recently, which broadens the central galaxy CLF in young halos and minimizes 
the difference among the central CLFs in halos of different ages.

The faint end of the CLF, where the dependence on halo age becomes much 
stronger, is dominated by satellite galaxies. 
The correlation of the number of satellite galaxies with the halo formation 
time reflects the balance between the accretion and destruction processes
\citep{Gao04,Zentner05}. We find that, at a fixed mass, halos that formed earlier 
start to accrete satellites (subhalos) earlier. Then these satellites have 
a longer time to experience the orbital decay caused by the dynamical friction and will 
more likely merge with the main central halos. Subhalos surviving at 
$z\sim 0$ are dominated by those accreted recently. Therefore, even though 
young halos accrete their mass later than old halos, they have much more 
{\it surviving} satellites.

Since the stellar masses in central and satellite galaxies have opposite
dependences on the halo formation time, we may expect that the total stellar 
mass in a halo of fixed mass is a weak function of the halo age. Defining
the total stellar mass in a halo as the sum for galaxies with 
$M_{\rm star}>10^9\msun$, we find that, in general, there are somewhat more 
stars in halos formed earlier, but the difference between 
the average stellar masses in the youngest and oldest halos is within a 
factor of 2.

\section{Discussion and conclusions}

In this Letter, we investigate the dependence of the galaxy contents
on the halo formation time using two galaxy formation models, one 
with a semi analytic method utilizing the realistic halo assembly history 
from $N$-body simulations and the other with a smoothed particle 
hydrodynamics simulation including radiative cooling, star formation,
and energy feedback from galactic winds. 
Even though our techniques for building merger trees as well as our
simulation methods are independent,
we confirm the dependence of halo clustering 
on the formation time reported by \citet{Gao05}. We study the CLF and CSMF 
of galaxies as a function of the halo formation time 
and find that galaxies inside halos 
have a strong correlation with the halo formation time.
In halos that form earlier, there are fewer satellite galaxies, and 
their central galaxies appear to be more luminous, which can be understood
through the early onset of star formation and the merging of satellites 
onto central galaxies. The dependence of the halo occupation number 
of satellite galaxies on the formation time in this study is similar to that 
of subhalos presented in \citet{Wechsler05} (also see \citealt{Zentner05}), 
which is not surprising since satellite galaxies reside in subhalos.

The HOD or CLF model assumes that the {\it statistical
distribution} of galaxies inside halos depends only on halo mass.
However, we note that this assumption does not mean 
that at fixed halo mass, the occupation number is independent of halo age. 
The correlation between the occupation number and the halo formation time exists 
even if the halo assembly history is based on the excursion set theory (e.g., 
\citealt{Zentner05}), and the distribution of the halo age at fixed mass could 
provide the scatter that goes into the probability distribution of the occupation 
number at fixed halo mass [e.g., $P(N|M)$ in the HOD model]. 
It is the dependence of halo clustering on halo age that makes the above 
correlation potentially important in interpreting galaxy clustering data.
Given the age-dependent halo clustering, the age dependence of HOD/CLF 
constitutes a key ingredient for a more accurate halo-based model of galaxy 
clustering. In view of this, a detailed study of the correlation between 
galaxy contents and halo formation time, like that performed in this 
Letter, is desired. 

Depending on the application, the current version of the HOD/CLF model without 
taking into account the non-Markovian nature of halo formation, can still 
give accurate descriptions (e.g., \citealt{Yoo05}). We reserve a detailed 
investigation for future work, to see to what extent it is adequate in 
different applications.

\acknowledgments 
We thank Volker Springel for his hydrodynamical recipes and for his useful comments
and David Weinberg for his helpful discussions.
The work in Shanghai was supported by the grants from
NSFC and from the Shanghai Key Projects in basic research (04JC14079
and 05XD14019).  Z. Z. acknowledges the support of NASA through Hubble
Fellowship grant HF-01181.01-A awarded by the Space Telescope Science
Institute, which is operated by the Association of Universities for
Research in Astronomy, Inc., for NASA, under contract NAS 5-26555. The
hydrodynamical simulations were run at the Shanghai Supercomputer Center.

\end{document}